\documentclass[12pt]{article}
\usepackage{amsmath,latexsym}
\usepackage{graphicx}
\begin{document}

\setlength{\unitlength}{1mm}

\title{Construction of 3D wormhole supported  by phantom energy}
 \author{\Large F.Rahaman$^*$, M.Kalam$^{\ddag}$,B.C.Bhui$^\dag$
and S.Chakraborty$^\dag$ }

\date{}
 \maketitle

 \begin{abstract}
               {\Large    In this article, we have found a series  solution of 3D Einstein
               equations describing a  wormhole for an inhomogeneous
               distribution of phantom energy. Here, we  assume equation of state is  linear but
               highly anistropic.  }

  \end{abstract}
  \footnotetext{\Large Pacs Nos :  04.20 Gz,04.50 + h, 04.20 Jb   
\\
 $*$Dept.of Mathematics, Jadavpur University, Kolkata-700 032,

 India

 E-Mail:farook\_rahaman@yahoo.com\\

$\ddag$Dept. of Phys. , Netaji Nagar College for Women, Regent Estate, Kolkata-700092, India.\\

  $\dag$Dept. of Maths., Meghnad Saha Institute of Technology,
                                           Kolkata-700150, India
}
    \mbox{} \hspace{.2in}


    \mbox{} \hspace{.2in}
\pagebreak

Pure gravity in (2+1) dimensions is fascinating in its own right.
In this case
 gravity does not propagate outside the sources i.e. no gravity out side matter.
 That means matter curves spacetime only locally. In other words, there is no gravitational waves.
 In (2+1) dimensional spacetime, Newtonian theory can not be obtained as a limit of Einstein's
 theory. Wormhole structure can not be obtained from Newtonian gravity. But it is welknown
 that Einstein's general theory of relativity admits wormhole structure in spacetime since
 according to Einstein's theory, the presence of matter twists the geometric fabric of spacetime.
 For this reason, why we consider ( 2+1) dimensions to discuss wormhole structure. Now,
wormholes are classical or quantum solutions for the gravitational
field equations describing a bridge between two asymptotic
manifolds. Classically, they can be interpreted as instantons
describing a tunneling between two distant regions. In a pioneer
work, Morris and Throne [1] have shown that the construction of
wormhole would require a very unusual form of stress energy
tensor. The matter that characterized the above  stress energy
tensor is known as exotic matter. This exotic i.e. hypothetical
matter can be of the following form either energy density of
matter $\rho < 0$ or  $\rho > 0$ but pressure $p< 0$. In 21st
century, this concept of negative energy is not pure fantasy, some
of its effects have been produced in the laboratory [2]. Recent
astronomical and cosmological observations indicate that the
Universe is undergoing
 a phase of accelerated expansion [3]. Theoretical Physicists  beleive that this acceleration
 is due to some unusual source of matter with positive energy density $\rho > 0$ and with negative
  pressure $p< 0$. This unsual matter source with the property, energy
density, $\rho > 0 $ but pressure $ p  < 0 $ is known as phantom
energy . Some how, if an advanced engineer could able to collect
this unusual source of matter i.e phantom energy, then it would be
possible to construct a wormhole. If wormhole could be constructed
, the faster than light travel would be possible in other words,
the time machine might be constructed. Now scientists are
interested to know how much negative energy is needed to construct
a wormhole. In 2003,  Visser , Kar
 and Dadhich [4] have proposed that wormholes could be
constructed with arbitrary small quantities of exotic matter. In
 recent past,  Delgaty et al [5] have studied traversable  wormhole  in (2+1) dimension
 with a cosmological constant. In this article, we are decided to provide a prescription of 3D
wormhole geometry by  using phantom energy as source. One could
imagine that an advanced engineer may use these results to
construct and sustain a traversable wormhole so that the
interstellar distances to be travelled in very short times.   And
we think, this might go forward our world.

As our target to provide a mathematical prescription of wormhole
geometry in (2+1) dimensions, we assume spherical symmetric metric
as

\begin{equation}
                ds^2 = - e^{2f(r)} dt^2 + \frac{1}{[1 - \frac{b(r)}{r}]}dr^2+r^2 d\phi^2
            \label{Eq1}
          \end{equation}
where,   $ r     \epsilon   (-\infty , +\infty) $ . \\

\pagebreak

To represent a wormhole , one must impose the following
conditions on the  metric (1) as [6] :  \\
1) The redshift function, $f(r)$ must be finite for all values of
$r$ . This means no horizon exists in the space time . \\
2) The shape function, $b(r)$ must obey the following conditions
at the throat $ r = r_0 $ : \linebreak $b(r_0) = r_0$ and
$b^\prime(r_0) < 1 $ [these are known as Flare-out conditions].\\
3) $\frac{b(r)}{r} < 1 $ for $ r >r_0 $ i.e. out of throat .\\
4)The space time is asymptotically flat i.e.$\frac{b(r)}{r}
\rightarrow 0 $ as $ \mid r \mid \rightarrow \infty $.\\

Using the Einstein field equations
 $G_{\mu\nu} = 8\pi T_{\mu\nu} $, in orthonormal reference frame
 ( with $ c = G = 1 $ ) , we obtain the following stress energy
scenario,

\begin{equation}
                8\pi\rho(r) =\frac{b^\prime r - b}{2 r^3}
                \label{Eq2}
          \end{equation}

\begin{equation}
                   8\pi p(r) = \frac{[ 1 - \frac{b}{r}]f^\prime}{r}
                \label{Eq3}
          \end{equation}

          \begin{equation}
               8\pi p_{tr}(r) =( 1 -
                \frac{b}{r}) [ f^{\prime\prime} - \frac{(b^\prime r - b )}{2r(r-b)}f^\prime
              + {f^\prime}^2 ]
                \label{Eq4}
          \end{equation}
where $\rho(r) $ is the energy density, $p(r)$ is the radial
pressure and $p_{tr}(r)$ is the transverse pressure. \\

Using the conservation of stress energy tensor $ T^{\mu\nu}_{;\nu}
= 0 $, one can obtain the following equation
\begin{equation}
                p^\prime  + f^\prime \rho + ( f^\prime +  \frac{1}{r})p -
                \frac{p_{tr}}{r} = 0
                \label{Eq5}
          \end{equation}

From now on , we assume that our source is characterized by the
 Phantom Energy with equation of state that contains a radial
 pressure
 \begin{equation}
               p =  - k \rho
                \label{Eq6}
          \end{equation}
we suppose also that pressures are highly anisotropic and
\begin{equation}
               p_{tr} = a  \rho
                \label{Eq7}
          \end{equation}

         \pagebreak

         From (5) by using (6) and (7), one can obtain
\begin{equation}
                \rho(r)e^{(1-\frac{1}{k})f} =\frac{\rho_0}{r^{(1+\frac{a}{k})}}
                \label{Eq10}
          \end{equation}
where $ \rho_{0} $ is an integration constant.\\

     Taking into account equations (2)-(8), we have the
following equation containing 'b' as
\begin{equation}( b^\prime r - b )^2 + ( r- b )[A(b^{\prime\prime}r^2- 3b^\prime
r+ 3b) + 2B( b^\prime r - b )] = 0
          \end{equation}
where $ \frac{1}{1-k} = A$ and $ B = \frac{k+a}{k(1-k)}$. \\

Now to investigate whether there exists physically meaningful
solutions consistent with the boundary requirements [ conditions
(1) to (4) ], we take a general functional form of $b(r)$ . We can
generally express it in the form
\begin{equation}
                b(r) = \Sigma_{n=1}^{\infty} b_n r^n +
                \Sigma_{m=0}^{\infty} a_m r^{-m}
                \label{Eq12}
          \end{equation}
since $\frac{b(r)}{r} \rightarrow 0 $ as $ r \rightarrow \infty $,
equation (12) is consistent only when all the $ b_n$'s in $b(r)$
vanish i.e.
\begin{equation}
                b(r) =  \Sigma_{m=0}^{\infty} \frac{a_m}{r^m}
                \label{Eq13}
          \end{equation}
Plugging this in equation (9) and matching the coefficients of
equal powers of $r$ from both sides , we get ,

$ 3A = 2B $ and $48A - 12 B = 36$ and these imply $ k =
\frac{1}{6}$ ,  $ a = \frac{1}{12}$.

Finally, one gets, the following form of b as
\begin{eqnarray*} \\&&
                b(r) =  a_0 - \frac{5a_0^2}{12r}+\frac{5a_0^3}{108r^2}
                +\frac{5a_0^4}{2976r^3}+\frac{587a_0^5}{964224r^4}
                + .......
                \label{Eq14}
         \end{eqnarray*}
Thus one gets, one parameter family of solutions.

 Now the expressions for
$\rho$ and f can be obtained as
\begin{equation}
                \rho =
                \frac{1}{16\pi}[  \frac{5a_0^2}{6r^4}-\frac{a_0}{r^3} -\frac{15a_0^3}{108r^5}
                -\frac{20a_0^4}{2976r^6}
                - .......]
            \label{Eq15}
          \end{equation}
\begin{figure}[htbp]
    \centering
        \includegraphics[scale=.8]{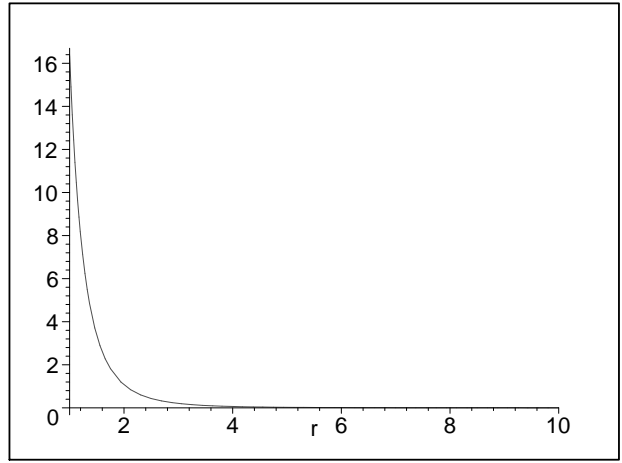}
        \caption{ Energy density  with respect to radial coordinate 'r' ( choosing  suitably the
        parameter) }
    \label{fig:1}
\end{figure}

\begin{equation}
                e^{2f} = [ \frac{( \frac{5a_0^2}{6r^{\frac{5}{2}}}-\frac{a_0}{r^{\frac{3}{2}}} -\frac{15a_0^3}{108r^{\frac{7}{2}}}
                -\frac{20a_0^4}{2976r^{\frac{9}{2}}}
                - .......)}{16\pi\rho_0}]^{\frac{2k}{1-k}}
            \label{Eq16}
          \end{equation}

\pagebreak

The throat of the wormhole occurs at $ r = r_0 $ where $ r_0 $
satisfies the equation
 \begin{equation}b(r) = r \end{equation}

Retaining a few terms, the graph of the function $F(r)=  b(r) -r $
indicates the point $r_0$ where $F(r)$ cuts the 'r' axis. From
the graph, one can also note that when $r>r_0 $, $F(r)< 0$ i.e.
$  b(r) -r < 0 $. This implies $ \frac{b(r) }{r} < 1 $ when
$r>r_0 $.

\begin{figure}[htbp]
    \centering
        \includegraphics[scale=.8]{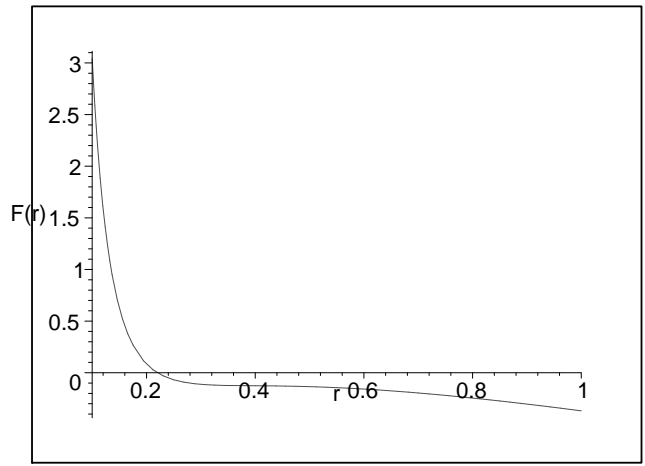}
        \caption{Throat occurs where $F(r)$ cuts 'r' axis}
   \label{fig:wh20}
\end{figure}

One can note that the redshift function $ f(r)$ always finite for
$  0 < r_0  \leq r  < \infty $ i.e. no horizon exists in the space
time. Thus our solution describing a static spherically symmetric
wormhole supported by the phantom energy.

\pagebreak

  According to Morris and Throne [1] , the 'r'
co-ordinate is
ill-behaved near the throat, but proper radial distance\\
\begin{equation}
 l(r) = \pm \int_{r_0^+}^r \frac{dr}{\sqrt{1-\frac{b(r)}{r}}}
            \label{Eq20}
          \end{equation}
 must be well behaved everywhere i.e. we must require that $ l(r)
 $is finite throughout the space-time . \\

 For our Model,
\begin{equation}
 l(r) = \pm \int_{r_0^+}^r
 \frac{dr}{\sqrt{1-\frac{1}{r}[a_0-\frac{5a_0^2}{12r}+\frac{5a_0^3}{108r^2}+....]}}
            \label{Eq21}
          \end{equation}
Though we can not find the explicit form of the integral but one
can see that the above integral is a convergent integral i.e.
proper length should be finite . \\

The axially symmetric embedded surface $ z = z(r)$ shaping the
Wormhole's spatial geometry is a solution of

\begin{equation}\label{Eq21}
 \frac{dz}{dr}=\pm \frac{1}{\sqrt{\displaystyle{\frac{r}{b(r)}}-1}}
 \end{equation}

  One can note from the definition of Wormhole that at   $ r= r_0 $
  (the wormhole throat) Eq.(17) is divergent i.e.  embedded surface is
   vertical there.

   The embeded surface   ( solution of eq.(17) ) in this case

      \begin{equation}\label{Eq22}
z = \pm \sqrt{a_0}[ 2\sqrt{r}- \frac{7}{12}a_0 r^{-\frac{1}{2}}
 - \frac{199}{648}a_0^2 r^{-\frac{3}{2}}-.....]
\end{equation}

One can see that embedding diagram of this wormhole (retaining a
few terms)
 in Fig-3.
 The surface of revolution of the curve about the vertical z axis makes
 the diagram complete (see  fig.4).

\begin{figure}[htbp]
    \centering
        \includegraphics[scale=.8]{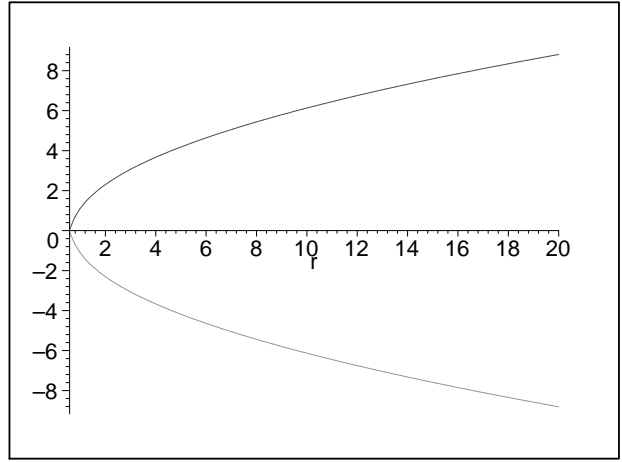}
        \caption{The embedding diagram of the wormhole ( choosing  suitably the
        parameter)  }
   \label{fig:3d-1}
\end{figure}

\begin{figure}[htbp]
    \centering
        \includegraphics[scale=.8]{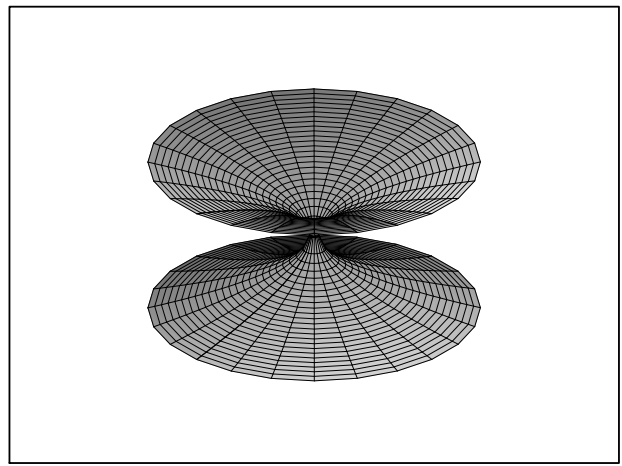}
    \caption{The full visualization of the surface generated by the rotation of the embedded
    curve (   retaining a
few terms ) about the vertical z axis }
    \label{fig:wormhole}
\end{figure}

\pagebreak

 In conclusion, our aim in this article has been provide a mathematical prescription for
 obtaining wormhole in 3D spacetime.  The source is realized by phantom energy.
One can also note that the equation of state is linear but highly
anisotropic. We see that shape function of our model satisfies all
conditions that are required  to represent a wormhole. The
resulting line element represents an  one parameter family of
geometries which contains wormholes. We note that as $ r
\rightarrow \infty $, the redshift function does not exist. Thus
our 3D wormhole characterized by phantom energy can not be
arbitrarily large. Also it may be assumed a 'cutoff' of the
stress energy tensor at a junction radius 'a', where the interior
wormhole metric will match to the exterior vacuum solution . We
end the article with the final remarks as if an advanced engineer
would able to collect sufficient amount of phantom energy, then
one can imagine that they should construct wormhole with the help
of several toy models [7] including this.

\pagebreak

        { \bf Acknowledgements }

        F.R is thankful to Jadavpur University and DST , Government of India for providing
          financial support. MK has been partially supported by
          UGC,
          Government of India under MRP scheme.  We are grateful to
          anonymous referee for pointing out the errors
          of the paper and for his
          constructive suggestions, which has led to a stronger result than the one in the
          original version.\\


\end{document}